# Hong-Ou-Mandel interference of two independent continuous-wave coherent photons

Heonoh Kim,[†] Danbi Kim,[†] Jiho Park, Han Seb Moon[*]

*Department of Physics, Pusan National University, Geumjeong-Gu, Busan, 46241, South Korea*
*Corresponding author: hsmoon@pusan.ac.kr*

**Interference between two completely independent photons lies at the heart of many photonic quantum information applications such as quantum repeaters, teleportation, and quantum key distribution. Here, we report the observation of Hong-Ou-Mandel (HOM) interference with two independent continuous-wave coherent light sources that are neither synchronized nor share any common reference. To prepare highly indistinguishable photons from two independent laser sources, we employ high-precision frequency-stabilization techniques using the $5S_{1/2}(F = 3)–5P_{1/2}(F' = 3)$ transition line of $^{85}$Rb atoms. We successfully observe a HOM interference fringe with two independent continuous-wave coherent photons originating from either the frequency-locked and frequency-modulated lasers. An interference fringe involving two-photon beating is also observed when the frequency difference between the two interfering photons is beyond the spectral bandwidth of the individual coherent photons. We carry out further experiments to verify the robustness of the source preparation regardless of the separation distance between the two independent photon sources.**

## 1. INTRODUCTION

The phenomenon of interference and its observation have been the focus of research in natural science and technology development for a long time. In particular, interferences of individual particles such as atoms [1], electrons [2], neutrons [3], molecules [4] and photons [5] have played an essential role in our understanding of the superposition principle and wave-particle duality in quantum mechanics. In this context, we begin with Dirac's famous statement on photon interference: "*Each photon then interferes only with itself. Interference between different photons never occurs*" [6]. Based on this idea, many experimental and theoretical studies have been undertaken to investigate the interference between two independent light sources [7-17]. In the early days, it was very difficult to observe the characteristic spatial/temporal fringe in first-order interference [7,8,12,14]. In addition, there were diverse scientific viewpoints on the critical conditions related to the observation of interference between two independent light sources and on the difference between the classical and quantum descriptions of the interference of light fields [7-14]. The essential requirement in the observation of the first-order interference fringe with two independent light beams, is to satisfy the critical condition that the measurement time must be significantly shorter than the reciprocal spectral bandwidth (or coherence time) of the involved light sources.

Meanwhile, since the early 1960s, Leonard Mandel's pioneering research on the interference between two independent light sources that are strongly attenuated to the single-photon level, has provided a starting point for new experimental perspectives on the interference of independent photons. Many of these experiments were performed utilizing photon counting and correlation techniques [15-18]. In this context, it must be noted that Mandel's studies are important because they demonstrated the interference effect between two independent photons even with sparsely localized photons. In such a case, it is not relevant if the measurement time is significantly longer than the coherence time. In particular, the first nonclassical interference experiment with correlation measurements involving two correlated photons from the spontaneous parametric down-conversion (SPDC) process [19] and intensity correlation experiments by Hanbury Brown and Twiss [20,21] have set the conceptual basis and framework for experimental quantum optics and quantum information science/technology.

The most interesting and important two-photon interference experiment employing two identical photons from SPDC was reported by Hong, Ou, and Mandel (the HOM effect) [22]. The physics underlying the HOM effect includes both the two-photon analogy of Dirac's statement on one-photon interference [6] and the quantum mechanical interpretation based on the interference between two alternative two-photon probability amplitudes [9,22,23]. Recently, observation of the HOM effect with identical (or indistinguishable) photons or entangled photons has played an important role in the practical implementation of photonic quantum information technologies [24-26]. Among these experiments, HOM experiment with two independent heralded photons [27-31] is of significant interest in terms of its application to quantum teleportation and entanglement swapping [32,33].

In the meanwhile, the first HOM experiments with low-coherence classical light source has been performed by Ou *et al.* for determining the second-order coherence time of a light beam based on intensity-correlation technique [34], and then this method has also been employed to measure the coherence time and pulse width of ultrafast pulses [35]. Recently, the two-photon interference experiments using classical light sources have been extensively performed by Liu et al. with two dissimilar photon sources [36, 37] and especially with two independent lasers [38, 39]. In addition, HOM experiments with two fully independent coherent light sources have also played a key role toward practical implementation of the measurement-device-independent quantum key distribution (MDI-QKD) protocol [40-44].

To date, most HOM experiments with two independent coherent light sources have been performed under pulse-mode operation [40-45] because the experiments have mainly focused on applying the QKD protocol. Recently, some experiments with the application of continuous-mode operation for spectroscopic and QKD applications have been demonstrated [46-48]. Indeed, there is no critical difference between pulse- and continuous-mode coherent light sources in observing the HOM-type interference effect except for the temporal restriction of the detected photons. The observation of HOM interference between two completely independent and autonomous photons can contribute to the realization of practical quantum applications.

In this paper, we experimentally demonstrate time-resolved HOM interference by employing two completely independent continuous-wave coherent photons (CWCPs) that are neither synchronized nor share any common reference. We used two external-cavity diode lasers (ECDLs) as narrowband coherent light sources and time-resolved measurements for two-photon coincidence counting. Furthermore, we employed two kinds of frequency-stabilization techniques using the spectra of $^{85}$Rb atoms to prepare indistinguishable photons with higher coherence from the two locally separated ECDLs. Our frequency-stabilization technique utilizing high-resolution atomic spectroscopy can aid in the realization of long-distance quantum communication employing coherent light sources; this is because the approach guarantees the reference-free and absolute preparation of two fully-independent coherent photons regardless of the separation distance between the light sources.

## 2. EXPERIMENTAL SETUP

Figure 1(a) shows the experimental setup utilized to realize time-resolved HOM interference with two independent narrowband CWCPs from two autonomously operated ECDLs that neither have an electrical/optical synchronization channel nor share any common reference. However, the two ECDLs are individually frequency-stabilized to the highly precise spectra of $^{85}$Rb atoms using error-signal locking obtained via saturated absorption spectroscopy (SAS) involving frequency modulation for ECDL1 and offset locking of polarization spectroscopy (PS) for ECDL2. Figures 1(b) and 1(c) show the SAS spectrum (blue curve) corresponding to ECDL1 and the PS spectrum (red curve) corresponding to ECDL2, respectively, for the $5S_{1/2}(F = 3)$–$5P_{1/2}(F' = 2,3)$ transition of $^{85}$Rb. The frequencies of ECDL1 and ECDL2 correspond to the optical frequency (384 228 115 MHz) of the $5S_{1/2}(F = 3)$–$5P_{1/2}(F' = 3)$ transition of $^{85}$Rb, as indicated by the blue and red arrows in Figs. 1(b) and 1(c), respectively.

In our experiment, the two ECDLs are independently operated and spatially separated, and therefore, the CWCP sources afford the advantages of their photons possessing an identical and high frequency stability arising from frequency stabilization with respect to the hyperfine atomic transition line. The two CWCPs are coupled with two single-mode optical fibers (SMFs) to eliminate spatial-mode mismatch and the two CWCPs are subsequently made incident on a 50:50 nonpolarizing beamsplitter (BS). The polarizations of the two input photons are controlled by half (H)- and quarter (Q)-wave plates positioned at the two SMF input ports. After passage through the BS, the superposed output photons are detected by two SMF-coupled single-photon detectors (SPDs) (SPCM-AQRH-13-FC, Excelitas technologies). The output signals from the two SPDs reach the time-correlated single-photon counter (TCSPC) for two-fold coincidence counting and the subsequent time-resolved measurement of the HOM interference of the two independent CWCPs [49,50].

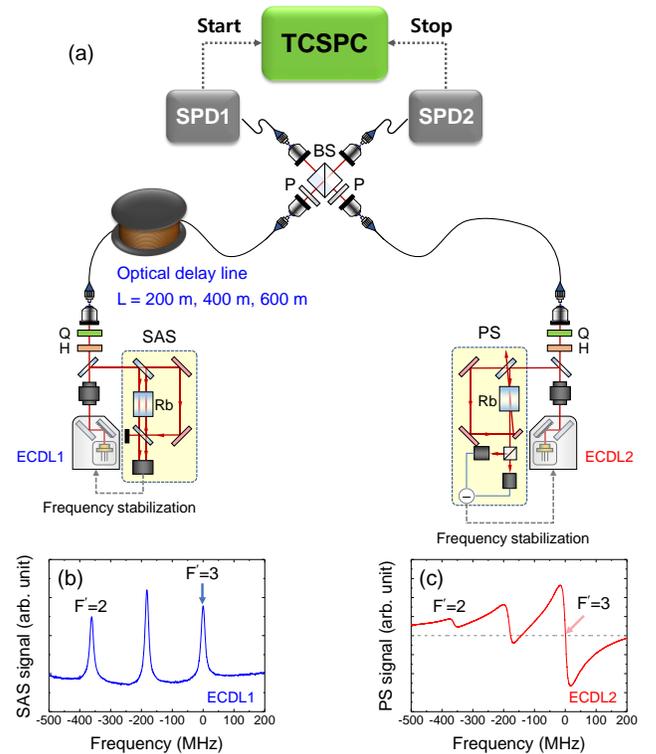

Fig. 1. Experimental setup for observation of Hong-Ou-Mandel (HOM) interference with two completely independent continuous-wave coherent photons (CWCPs). (a) Schematic of the experimental setup for HOM interference between two independent CWCPs (P: polarizer, H: half-wave plate, Q: quarter-wave plate, BS: beam splitter, SPDs, single-photon detectors, TCSPC: Time-correlated single-photon counter, Rb: atomic vapor cell of $^{85}$Rb, SAS: saturated absorption spectroscopy, PS: polarization spectroscopy). (b) SAS spectrum (blue curve) of ECDL1 and (c) PS spectrum (red curve) of ECDL2 for the $5S_{1/2}(F = 3)$–$5P_{1/2}(F' = 2,3)$ transition of $^{85}$Rb.

## 3. RESULTS

To observe a stable HOM interference fringe with two independent CWCPs, it is necessary to generate indistinguishable photons from two independent laser sources. Such indistinguishable independent CWCPs must exhibit identical spectral, spatial, and polarization modes at the two input ports of the BS. In particular, it is important to stably and accurately control the spectral properties of the two CWCPs via the application of a high-resolution frequency-stabilization technique. Therefore, in our work, we measured the mutual spectrum between the two coherent light sources by using the beat signal between the two ECDLs. With the setup shown in Fig. 1(a), when the two ECDLs are operated in the free-running mode, the width of the beat spectrum is measured to be ~1 MHz, corresponding to the spectral width of our narrowband coherent light sources. When ECDL1 is frequency-stabilized by error-signal locking of the modulated SAS signal, the spectral shape becomes nearly rectangular (solid curve) with a width of 5.2 MHz, as shown in Fig. 2(a). The square shape is due to the modulation frequency of 1 kHz and modulation width of ~4 MHz. Figure 2(b) shows the spectral density spectrum of the frequency-stabilized ECDL2, we can observe a nearly Lorentzian shape (solid curve) with a spectral width of 2.2 MHz. The spectral bandwidth is more broadened than that in the case of the free-running ECDL because of the PS offset locking.

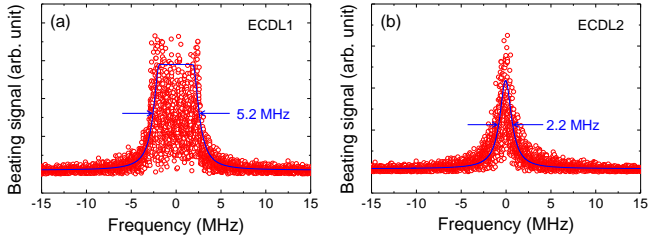

Fig. 2. Spectral properties of two independent continuous-wave coherent photons. Spectral density spectrum of (a) frequency-stabilized ECDL1 upon error-signal locking of the SAS with frequency modulation, and (b) frequency-stabilized ECDL2 obtained by offset locking of the PS.

When the two independent CWCPs with the spectral properties shown in Fig. 2 are incident on the two input ports of the BS, the coincidence counting probability as a function of time delay ($\Delta T$) can be expressed as [47, 51]

$$P_{\text{coin.}}(\Delta T) = 1 - V\Gamma_{12}(\Delta T)\cos(\Delta\omega\Delta T), \quad (1)$$

where $V$ represents the HOM-fringe visibility, which is limited to 0.5 for classical light sources, $\Gamma_{12}(\Delta T)$ the mutual coherence function between the two input CWCPs, and $\Delta\omega$ the center-frequency difference of the two CWCPs. The HOM fringe shape and width are determined by convolving the two spectral properties shown in Fig. 2. Figure 3 shows the measured time-resolved HOM interference fringe with the two fully-independent and frequency-stabilized CWCPs prepared with the spectral properties shown in Fig. 2. The horizontal axis represents the detection-time delay between the two SPDs, where the time-bin is set to 0.5 ns. The coincidence counting rates accumulated over 200 s are normalized to 1 to ensure a significantly longer time delay than the mutual coherence time. From the theoretical curve fitting of the experimental result, the fringe visibility and width are estimated to be 0.432 ± 0.002 and 296 ± 2 ns, respectively. The expected fringe width can be estimated from the average spectral bandwidth of 3.1 MHz, which corresponds to a time delay of 323 ns.

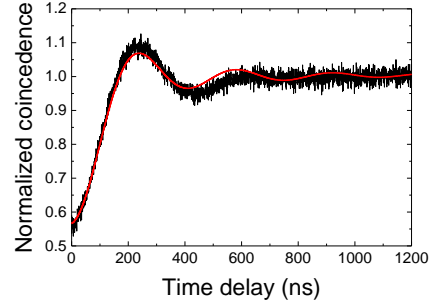

Fig. 3. Hong-Ou-Mandel interference fringe with two independent continuous-wave coherent photons. Experimental result and theoretical curve fitting under the conditions of a bandwidth of 3.3 MHz and $\Delta\omega = 0$.

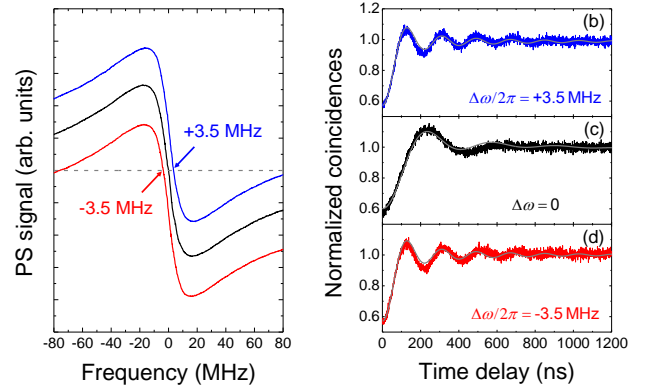

Fig. 4. Two-photon beat fringes for different offset frequencies. (a) PS spectra (blue: $\Delta\omega/2\pi$ = +3.5 MHz, black: $\Delta\omega = 0$, and red: $\Delta\omega/2\pi = -3.5$ MHz) of ECDL2 frequency-stabilized to the $5S_{1/2}(F = 3)$–$5P_{1/2}(F' = 3)$ transition of $^{85}$Rb; the horizontal dashed line (gray) indicates the DC-offset reference. (b-d) The HOM-type two-photon beating fringes are measured for different offset frequencies.

Next, we changed the optical frequency of ECLD2 by adjusting the offset of the PS spectrum, as shown in Fig. 4(a). In the figure, the intersection points of the dispersive-like PS spectra and the gray dashed line denote the frequency-locking points for stabilization. The respective intersection points of the blue, black, and red curves in Fig. 4(a) correspond to $\Delta\omega/2\pi$ = +3.5 MHz, 0, and -3.5 MHz, respectively. Figures 4(b)-4(d) show the measured HOM interference fringes including two-photon beating corresponding to the two cases shown in Fig. 4(a). Although the center-frequency difference is larger than the spectral bandwidth of the individual coherent light sources, the fringe visibility is not degraded. However, the theoretical fit does not have exactly the same shapes with experimental data due to a rather slowly modulated frequency to achieve the error-signal locking of the modulated SAS signal as shown in Fig. 2(a). This result indicates that the two independent CWCPs can be stably prepared by employing local and remote frequency-stabilization with reference to the high-precision atomic transition line, which is useful for obtaining stable HOM interference regardless of the separation distance of the light sources.

We conducted a further experiment to verify that the time-resolved measurement of the HOM-type two-photon interference of the two fully independent and narrowband CWCPs is not affected by any temporal delay between the two interfering photons at the BS input ports. We repeated the measurements under the same experimental conditions as those applied for obtaining the HOM interference fringe shown in Fig. 3. In the case wherein two independent CWCPs are used to observe the HOM interference, the relative phase between the two photons is completely random. However, if we employ the time-resolved measurement technique to observe the HOM interference, there is no limit to the optical path-length difference from the light sources to the BS. This is because the mutual coherence in this measurement is defined by the transient first-order degree of coherence of the superposed BS output photons conditioned by the detection of the single photons from either of the two BS input ports. Figure 5 shows the experimental results for three optical delay lines (ODLs) generated by including SMF spools (200 m, 400 m, and 600 m) in the path of ECDL1. Although the input time delay obtained with the use of the ODL is considerably longer than the coherence length of the individual CWCPs, the HOM interference is not degraded, and the visibility also does not change.

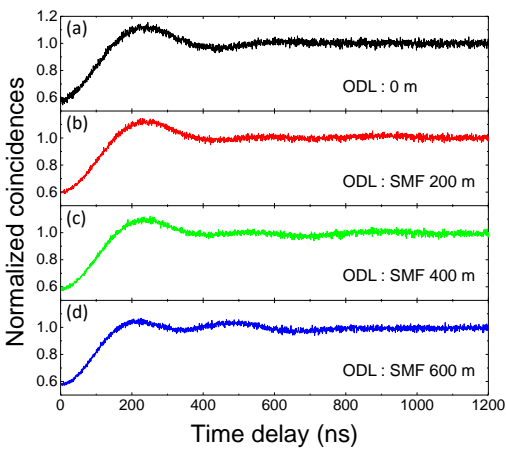

Fig. 5. Hong-Ou-Mandel interference fringes with arbitrary time delay between two independent CWCP sources. Three single-mode fiber (SMF) spools are employed as relative optical delay lines (ODLs). (a) SMF = 0 m (black curve), (b) SMF = 200 m (red curve), (c) SMF = 400 m (green curve), and (d) SMF = 600 m (blue curve).

## 4. CONCLUSION

In conclusion, we successfully demonstrated HOM interference with weak coherent photons from two completely independent continuous-wave lasers without employing any synchronization mechanism between them. The indistinguishable CWCPs were stably generated with the use of high-resolution frequency stabilization that utilized frequency locking to the hyperfine transition line of the $^{85}$Rb atom. Our CWCP sources exhibit the advantages of possessing a universal identity and high frequency stability, which guarantees the reference-free or absolute preparation of two independent and highly coherent photons regardless of the separation distance between the light sources. We believe that our approach can provide an efficient tool for the spectral characterization of light sources based on the two-photon interference of weak coherent light sources prepared by employing well-established high-resolution atomic spectroscopy. Moreover, the approach can be used to prepare highly indistinguishable coherent photons for the practical implementation of quantum communication systems.

**Funding.** National Research Foundation of Korea (NRF) (2018R1A2A1A19019181).

**Disclosures**. The authors declare no conflicts of interest.

†These authors contributed equally to this work.